\documentclass[11pt]{article}
\usepackage{graphicx}
\begin{document}
\begin{center}
{\bf Structure effects of the two protons - two neutrons correlations\footnote{In {\it Contributions at the Int. Summer School "Recent Advances in Nuclear Structure", Predeal, Romania - 1990}, Edited by D. Bucurescu, Gh. C\u ata-Danil and N. V. Zamfir, Rev. Roum. Phys., Tome {\bf 37}, N$^o$ 2, Bucarest (1992), p. 107-113.} }\\[.5cm]
M. Grigorescu \\[1cm]
\end{center}
\noindent
$\underline{~~~~~~~~~~~~~~~~~~~~~~~~~~~~~~~~~~~~~~~~~~~~~~~~~~~~~~~~
~~~~~~~~~~~~~~~~~~~~~~~~~~~~~~~~~~~~~~~~~~}$ \\[.1cm]
{\bf Abstract}:
{\small This work presents the proton-neutron 2 particles - 2 holes ground state correlations arising from the four-particle interaction. As a direct test for such a mean-field, the low-energy monopole resonances will be shortly discussed. } \\
$\underline{~~~~~~~~~~~~~~~~~~~~~~~~~~~~~~~~~~~~~~~~~~~~~~~~~~~~~~~~
~~~~~~~~~~~~~~~~~~~~~~~~~~~~~~~~~~~~~~~~~~}$ \\[.5cm]

\newpage
\section{ Introduction}  
The $\alpha$-particle type correlations between protons and neutrons have been investigated after the early success of the BCS theory in nuclei \cite{1}, in connection both with the charge-independent treatment of the two-body force \cite{2}, as with the four-particle interaction \cite{3}. In the first case, the correlations were accounted explicitly by the extension of the BCS trial states as to include the quadruples, while in the second case the correlated ground state was obtained self-consistently, as eigenfunction of the total Hamiltonian linearized in the four-body term. Intending to obtain a quadruple mean-field, the linearization was restricted to the proton-neutron (p-n) 2 particles-2 particles channel, although the    
linearization in the p-n 2 particles-2 holes (2p-2h) channel is also non-trivial. \\ \indent
As it was shown previously \cite{4}, a stable p-n, 2p-2h mean-field, may become energetically favoured, leading as a ground-state collective effect to the symmetry energy of the even-even systems of nucleons lying on the same degenerate level. Moreover, this mean-field by its Josephson-like structure leads to isovector monopole excitations of low energy, appearing only in superfluid nuclei, corresponding to a pair exchange between the neutron and proton superfluids. \\ \indent
In this work, the p-n, 2p-2h mean-field will be considered within a boson approximation, extending the previous treatment of a single $j=3/2$ level. Also, the electroexcitation of the isovector monopole states will be shortly discussed.  
\section{Mean-field approximations for the four-particle interaction} 
As the mean-field approximation requires first a separable interaction, an important reference is the quadrupling Hamiltonian presented in \cite{5}
\begin{equation}
H_{Q0} = - \frac{G_Q}{4} Q_{00}^\dagger Q_{00}
\end{equation}
with 
$$
Q_{00}^\dagger = 2 \sqrt{3} [{\cal P}^\dagger {\cal P}^\dagger]_{00}~~.
$$

Here $[{\cal P}^\dagger {\cal P}^\dagger]_{TT_0}$ denotes the isospin Clebsch-Gordan coupling and ${\cal P}^\dagger_\mu$, $\mu=-1,0,1$, is an isovector with the components defined by the proton-proton, proton-neutron and neutron-neutron pairing operators $P_+$, $N_+$, $R_+$, 
$$
{\cal P}^\dagger_{-1}=P_+~~,~~{\cal P}^\dagger_0= \frac{1}{\sqrt{2}} R_+~~,~~ {\cal P}^\dagger_1= N_+
$$ 
$$
P_+  =  \frac{1}{2} \sum_{a,m} s_{am} c^\dagger_{pam} c^\dagger_{pa-m}~~,~~ P_+^\dagger \equiv P_-~,a \equiv (n,l,j)~~,s_{am}=(-1)^{j-m} 
$$ 
$$
R_+  =  \frac{1}{2} \sum_{a,m} s_{am} c^\dagger_{nam} c^\dagger_{pa-m}~~,~~ R_+^\dagger \equiv R_- 
$$
\begin{equation}
N_+  =  \frac{1}{2} \sum_{a,m} s_{am} c^\dagger_{nam} c^\dagger_{na-m} ~~,~~N_+^\dagger \equiv N_- ~~.
\end{equation}
For nuclei with a large neutron excess the p-n pairing can be neglected, and the main component of $H_{Q0}$ becomes 
\begin{equation}
H' = - G' P_+N_+P_-N_-~,~~G'=4G_Q~~.
\end{equation}
This restricted four-particle interaction was considered with success recently \cite{6} to account for the observed correlations between the proton and neutron superfluidity in medium and heavy nuclei. \\ \indent
By contrast to the quadrupling term $H_{Q0}$, which is clearly separable in the 2p-2p channel, $H'$ may be equally well separated in the p-n, 2p-2p or 2p-2h channels. Excepting for the terms renormalizing the usual pairing interaction, the corresponding mean-fields are obtained when the self-consistent calculation leads to non-vanishing ground-state expectation values $\Delta_4 = G' <P_+N_+>$, $\delta_4= G' <P_+N_->$. \\ \indent
Although straightforward, this approach to the p-n, 2p-2h correlations has a weak point, namely to define both mean-field parameters $\Delta_4$, $\delta_4$ using the same interaction constant $G'$. Thus, it appears necessary to associate the p-n, 2p-2h mean-field, to a four-body interaction which is different from quadrupling, isoscalar and separable in the 2p-2h channel. A simple choice, satisfying these requirements and becoming similar to $H'$ when the proton-neutron pairing is neglected, is the isospin quadrupole interaction 
\begin{equation}
H_{Q2} = - G_4 \sum_{\mu=-2}^2 Q_{2 \mu}^\dagger Q_{2 \mu}~~,
\end{equation}
with 
$$
Q_{2 \mu}^\dagger = [{\cal P}^\dagger {\cal P}]_{2 \mu}~~.
$$
Its interaction constant $G_4$ appears as an additional parameter, which should be determined from experiment. 
\section{ The strength of the 2p-2h mean-field} 
The first estimates of the constant $G_4$ and of the mean-field parameter $\delta= 2G_4 <P_+N_->$ were obtained using the mass difference between the even-even members of the $T=2$ isomultiplet at $A=36$, within the approximation of a single $j=3/2$ level \cite{4}. Also, the quadrupling and the proton-neutron pairing terms were neglected, reducing the linearized Hamiltonian to
\begin{equation}
H_L= \epsilon (\hat{N}_p + \hat{N}_n) - G(P_+P_- + N_+N_-) - \delta (P_+N_- + P_-N_+) - \omega \hat{T}_0 ~~.
\end{equation}
Here $\hat{N}_p, \hat{N}_n$ are the particle number operators for protons and neutrons, $\hat{T}_0= (\hat{N}_n - \hat{N}_p)/2$ and $- \omega \hat{T}_0$ is a "cranking" term restoring in average the isospin symmetry broken by the p-n, 2p-2h mean-field. The parameter $\omega$ is fixed by the constraint 
\begin{equation}
\langle g_\omega \vert \hat{T}_0 \vert g_\omega \rangle = \frac{N-Z}{2}~~,
\end{equation}
with $\vert g_\omega \rangle$ the $\omega$-dependent ground-state of $H_L$. \\ \indent
At $\omega=0$, ($N=Z$), the self-consistency condition gives a mean-field parameter    
$$
\delta_0 = 2 G_4 \langle g_0 \vert P_+ N_- \vert g_0 \rangle = \sqrt{8 G_4^2 - G^2/8}~~,
$$ 
and the occurrence of the stable p-n, 2p-2h mean-field, is expected when $G_4 > G/8$. \\ \indent
For $\omega \ne 0$, but small, the ground-state energy $\langle g_\omega \vert H_L+ \omega \hat{T}_0 \vert g_\omega \rangle$ increases with respect to its value at $\omega =0$ by a term 
${\cal E}^\omega_g =(N-Z)^2/8J^0_g$,
$$
J^0_g = \frac{ 32 \delta_0^2}{\sqrt{G^2+8 \delta_0^2}(\sqrt{G^2+8 \delta_0^2}
+ G)^2}~~, 
$$       
having the same form as the symmetry energy $W_{sym}(A, N-Z)=k_w (N-Z)^2$, $k_w=28.1/A$ MeV, from the Weisz\"acker mass formula. \\ \indent
Observing that the Coulomb corrected ground-state energy differences between $^{36}$S or $^{36}$Ca and $^{36}$Ar are practically the same as $W_{sym}(36, 4)$, for the first estimate it has appeared natural to identify $k_W$ with $1/J^0_g$. \\ \indent
In fact, by considering a 4-degenerate $j=3/2$ level, $\omega$ tends to infinity when $N-Z= \pm 4$, and asymptotically  ${\cal E}^\omega_g$ takes the value ${\cal E}^\infty_g = G + \sqrt{G^2+8 \delta_0^2}$. This saturation, appearing due to the low degeneracy of the level $j=3/2$, and expected to disappear in a realistic calculation, excludes a direct comparison between  ${\cal E}^\infty_g$ and $W_{sym}(36, 4)$, but also it makes the infinitesimal comparison less relevant. So, to get a better idea about the range of $G_4$ and $\delta_0$, it becomes of interest to investigate the one-level system in the limit of large degeneracy. In this limit, when the level degeneracy, denoted $2 \Omega$, is high relatively to the number of valence particles (protons or neutrons), the operators $P_+,N_+,\hat{N}_p,\hat{N}_n$ can be represented using the boson operators $b_2,b_3,b_2^\dagger,b_3^\dagger$ as
\begin{equation}
P_+= \sqrt{\Omega} b_2^\dagger~~,~~N_+= \sqrt{\Omega} b_3^\dagger~~,~~ P_-= \sqrt{\Omega} b_2~~,~~N_-= \sqrt{\Omega} b_3   
\end{equation}
$$
\hat{N}_p = 2 b_2^\dagger b_2 ~~,~~\hat{N}_n = 2 b_3^\dagger b_3~~,
$$
and $H_L$ can be approximated by the Hamiltonian $H_L^B$ of two coupled harmonic oscillators
\begin{equation}
H_L^B= \omega_2 b^\dagger_2 b_2 +\omega_3 b^\dagger_3 b_3 -  \delta \Omega (b^\dagger_2 b_3+ b^\dagger_3 b_2) ~~,
\end{equation}
$$
\omega_2 = 2 \epsilon - G \Omega + \omega~~,~~ \omega_3 = 2 \epsilon - G \Omega - \omega~~.
$$
This Hamiltonian can be written as a sum of two independent boson terms 
\begin{equation}
H_L^B= \Omega_2 B^\dagger_2 B_2 +\Omega_3 B^\dagger_3 B_3~~, 
\end{equation}
$$
\Omega_2 = 2 \epsilon - G \Omega + \omega / \cos 2 \rho ~~,~~ \Omega_3 = 2 \epsilon - G \Omega - \omega / \cos 2 \rho~~,
$$
$$ \cos 2 \rho = \vert \omega \vert / \sqrt{ \delta^2 \Omega^2 + \omega^2} ~~,$$
where the operators $B^\dagger_2,B^\dagger_3$ are related to $b^\dagger_2,b^\dagger_3$ by the unitary transformation $U= \exp[ \rho (b^\dagger_2b_3 - b_2b^\dagger_3)]$,  
\begin{equation}
B^\dagger_2 = U b^\dagger_2 U^{-1} = \cos \rho~ b^\dagger_2 - \sin \rho ~b^\dagger_3~~,
\end{equation}
$$
B^\dagger_3 = U b^\dagger_3 U^{-1} = \sin \rho~ b^\dagger_2 + \cos \rho ~b^\dagger_3~~.
$$
If $\vert n_2, n_3 \rangle_\omega$, ($B^\dagger_k B_k \vert n_2, n_3 \rangle_\omega = n_k \vert n_2, n_3 \rangle_\omega $, $k=2,3$), are the eigenstates of $H^B_L$, it appears clearly that a four-particle system may be found in one of the three eigenstates $\vert 2, 0 \rangle_\omega$,
$\vert 1,1 \rangle_\omega$, $\vert 0,2 \rangle_\omega$. In general, none is the ground-state of $H^B_L$, but the one denoted $\vert 2 \Omega_m \rangle_\omega$,  having the lowest energy $E_g = 2 \Omega_m$, $\Omega_m = \min (\Omega_2,\Omega_3) =2 \epsilon - G \Omega - \sqrt{ \delta^2 \Omega^2 + \omega^2}$, will correspond, for appropriate values of $\omega$, to the ground-states 
of the even-even isobars having four valence particles. Similarly, for a total number of $A_v$ valence particles, $A_v \ll 4 \Omega$, the ground-state is represented by $\vert n_v \Omega_m \rangle_\omega$, $n_v=A_v/2$. \\ \indent
To achieve self-consistency, it is necessary to define each state $\vert n_2, n_3 \rangle_\omega$ using for $\delta$ the value denoted $\delta_\omega$, given by the equation
\begin{equation}
\delta_\omega = 2 \Omega G_4 \cdot ~_\omega\langle  n_2, n_3 \vert b^\dagger_2b_3 \vert n_2, n_3 \rangle_\omega~~.
\end{equation}
For the state $\vert n_v \Omega_m \rangle_\omega$  this leads to a mean-field parameter
\begin{equation}
\delta_\omega =  \sqrt{ (A_v \Omega G_4/2)^2 - (\omega / \Omega)^2}~~,
\end{equation}   
decreasing from the value $\delta_0 = A_v \Omega G_4/2$ in the $N=Z$ nucleus, to zero in the $T_0 = \pm A_v /2$ nuclei. Defined as previously, ${\cal E}^\omega_g$ now has the form 
$$ {\cal E}^\omega_g= \omega^2 / (\Omega^2 G_4)~~, $$
with $\omega = \Omega^2 G_4 (N-Z)/2$. This form is preserved in the whole range of $\omega$, and consequently the comparison with $W_{sym}(A, N-Z)$ gives a well-defined result: $\delta_0 = 2 A_v k_W / \Omega$, $G_4= 4 k_W / \Omega^2$. Assuming the effective value of $\Omega$ at the Fermi level to be roughly $\Omega_{eff} \sim \sqrt{A}$, than $G_4 \sim 112/A^2 $ MeV. For $^{36}$Ar this means $G_4 \sim 0.08$ MeV, in agreement with the previous estimate \cite{4}, but for $\delta_0$ a value about ten times larger is obtained.   
\section{ The electroexcitation of the low-energy monopole resonances} 
Besides the contribution to the symmetry energy, a direct indication about the presence of a p-n, 2p-2h mean-field could be the observation of low-energy monopole resonances \cite{4}. By contrast to the isovector breathing oscillations observed recently in charge-exchange reactions on a wide range of nuclei \cite{7} at an excitation energy of $170/A^{1/3}$ MeV, the oscillations due to the p-n, 2p-2h mean-field  are expected only in superfluid nuclei (when $\Delta = G \langle P_+ \rangle \approx G \langle N_+ \rangle \ne 0$), and at a lower energy,
$\omega_v = 8 \Delta\sqrt{k_W \delta_0} /G$. This low-energy oscillation can be pictured in the "coordinate" space as an opposite oscillation of the proton and neutron BCS phase angles, or in the "momentum" space as an oscillation in $T_0$. Thus, it is natural to investigate its excitation by the $T_0$-dependent monopole operator from the electron scattering cross-section.  \\ \indent
In the relativistic limit, when the electron mass can be neglected, the monopole term from the electroexcitation cross-section takes the form \cite{8} 
\begin{equation}
\frac{ d \sigma}{d \Omega } = 16 \pi \alpha^2 f_{rec} B(C0,q) \frac{   k_f^2}{q^4}  \cos^2 \frac{\theta}{2}
\end{equation}  
where $\alpha =1/137$, ${\bf k}_i$, (${\bf k}_f$), is the initial, (final), momentum of the electron, $ f_{rec} = M/(M+ 2k_i \sin^2 \theta /2)$,   
$M$ is the mass of the nucleus, ${\bf q}= {\bf k}_i-{\bf k}_f$, and $B(C0,q)$ is the transition strength for the operator 
\begin{equation}
{\cal M}_0 = \frac{1}{ 4 \sqrt{\pi}} \sum_{i=1}^A \int d^3 r (1+\tau^i_3) \delta( {\bf r} - {\bf r}_i) j_0 (qr) 
\end{equation}
from the $0^+$ ground-state $\vert 0^+, gd \rangle$ to the low-energy monopole state 
$\vert 0^+, \omega_v \rangle$, 
\begin{equation}
B(C0,q) = \vert \langle 0^+, \omega_v \vert {\cal M}_0 \vert 0^+, gd \rangle \vert^2~~. \label{bc0}
\end{equation}
Within the random phase approximation (RPA) for a single, half-filled, $2 \Omega$-degenerate level \cite{9}, the states $\vert 0^+, gd \rangle$ and $\vert 0^+, \omega_v \rangle$ are given by the equations 
\begin{equation}
{\cal Q} \vert 0^+, gd \rangle =0
\end{equation}
and 
\begin{equation}
{\cal Q}^\dagger \vert 0^+, gd \rangle = \vert 0^+, \omega_v \rangle
\end{equation}
with
\begin{equation}
{\cal Q}^\dagger = \frac{1}{2 \Omega} \sqrt{\frac{\omega_v}{\delta_0}} \lbrack \frac{ \delta_0 \Omega}{\omega_v} (P_+-P_- -N_++N_-) - \hat{T}_0 \rbrack~~.
\end{equation}
As it was shown by the semiclassical approach \cite{4}, the hermitian term $\hat{T}_0$ accounts for the BCS phase angles vibration, while the remaining nonhermitian term accounts for the associated oscillation of the gap parameters. \\ \indent
Writing ${\cal Q}^\dagger$ in terms of quasiparticles relatively to the BCS ground-state $\Psi_g$,
$$
\Psi_g = U_0 \vert 0 \rangle~~,~~U_0= \exp[ \pi (P_++N_+-P_--N_-)/4]
$$
as 
\begin{equation}
{\cal Q}^\dagger = \sum_{\tau=p,n} (X_\tau S^\dagger_\tau -Y^*_\tau S_\tau) ~~,
\end{equation}
\begin{equation}
S^\dagger_p = U_0 P_+ U^{-1}_0 ~~,~~S^\dagger_n = U_0 N_+ U^{-1}_0~~,~~
S_p= (S_p^\dagger)^\dagger~~,~~S_n= (S_n^\dagger)^\dagger~~,
\end{equation}
it can be shown that the state $\vert 0^+, gd \rangle$ exists only if $\Omega$ is even, and in this case is a product between a proton and a neutron function: $\vert 0^+, gd \rangle = \vert RPA_p \rangle\vert RPA_n \rangle$. The functions $\vert RPA_\tau \rangle$, $\tau =p,n$, appearing here are given, up to a normalization constant, by
\begin{equation}
\vert RPA_\tau \rangle \sim \sum_{k=0}^{\Omega /2} \frac{(\Omega - 2 k)!}{\Omega ! k!} \frac{ (\Omega/2 -k)!}{(\Omega /2)!} ( \frac{Y_\tau}{X_\tau } )^k (S_\tau^\dagger)^{2k} \Psi_g~~,
\end{equation}
containing an even number of quasiparticle pairs. \\ \indent
For small scattering angles ($\theta \sim 0$), the momentum transfer $q$ is also small, and using the expansion $j_0 (qr) \approx 1 - (qr)^2/6$, the first non-vanishing term in (\ref{bc0}) is     \begin{equation}
B(C0,q) \vert_{q \rightarrow 0} = \frac{1}{16 \pi}  \vert \langle 0^+, gd \vert [{\cal Q}, \hat{T}_0] \vert 0^+, gd \rangle \vert^2~~.
\end{equation}
This term is independent of $q$, as it should be for a "true monopole", by contrast to the first non-vanishing term for other monopole transitions, having the $q$-dependence of the quadrupole. 
\\ \indent
Approximating  in  $B(C0,q)$ the state  $\vert 0^+, gd \rangle$ by the uncorrelated ground-state $\Psi_g$, the scattering cross-section at low momentum transfer becomes 
\begin{equation}
\frac{ d \sigma}{d \Omega } \vert_{q \rightarrow 0} = 4 \pi \sigma_M B(C0,q) \vert_{q \rightarrow 0} ~~,
\end{equation}  
where   
\begin{equation}
B(C0,q) \vert_{q \rightarrow 0} = \frac{1}{4 \pi} \frac{\delta_0}{\omega_v} (\frac{\Delta}{G})^2~~,
\end{equation}
and 
\begin{equation}
\sigma_M = \frac{\alpha^2}{4 k^2_i} \frac{ \cos^2 \theta /2 }{\sin^4 \theta / 2} 
\end{equation}
is the Mott cross-section\footnote{e.g. in A. J. Buchmann, Charge form factors and nucleon shape,  arXiv:0712.4270.}.

\end{document}